\documentclass{elsart}

\usepackage{graphics}
\begin{document}
\begin{frontmatter}

\title{Jamming and Stress Propagation in Particulate Matter}

\author{M.~E. Cates, J.~P. Wittmer}
\address{ Dept. of Physics and Astronomy, University of Edinburgh\\
JCMB King's Buildings, Mayfield Road, Edinburgh EH9 3JZ, GB.}
\author{J.-P. Bouchaud, P. Claudin}
\address{Service de Physique de l'Etat Condens\'e,
CEA\\ Ormes des Merisiers, 91191 Gif-sur-Yvette Cedex,
France.}
\thanks{}
\begin{abstract}
We present simple models of particulate materials whose mechanical
integrity arises from a jamming process. We argue that
such media are generically ``fragile", that is, they are unable to support
certain types of incremental loading without plastic rearrangement.
In such models, fragility is naturally linked to the marginal stability of
force chain networks (granular skeletons) within the material. Fragile matter
exhibits novel mechanical responses that may be relevant to both
jammed colloids and cohesionless assemblies of poured, rigid grains.

\end{abstract}
\end{frontmatter}

\section{Introduction}\label{sec:Intro}

In this paper, we consider the relation between jamming (a
kinetic process) and the laws of static mechanical equilibrium of
particulate media. We describe first a simple model of jamming in
a colloid, sheared between parallel plates.  We assume that force chains
(linear arrays of rigid particles in contact) develop along the major
stress compression axis and span the sample. The resulting jammed state
can support a shear stress indefinitely, but does this by virtue of a
self-organized internal structure (the force chain array) which has
developed in direct response to the applied load itself. If a different
load is now applied (e.g. if the material is sheared in some different
direction), the force chains cannot sustain the load but must flow and
rejam in a different configuration. The model thus provides a simple
example of a ``fragile" material -- one which cannot support certain
types of infinitesimal stress increment without plastic reorganisation.

The continuum mechanics of fragile materials is very different from
conventional elastic or elastoplastic theory. We propose that
fragility may be generic in jammed systems, and argue further that a pile of
cohesionless poured sand can be viewed as jammed in the required sense: its
mechanical integrity results purely from the applied load (gravity). Some
simple models that we have recently developed for poured sand
\cite{nature,bcccargese,staticavalanche,PRLroysoc}, such as the
fixed principal axes ({\sc fpa}) model, indeed display fragile behaviour; in
fact they correspond to assuming a particular simplified geometry for the
granular skeleton of force chains.

\begin{figure}
\centerline{
\scalebox{.6}{\includegraphics{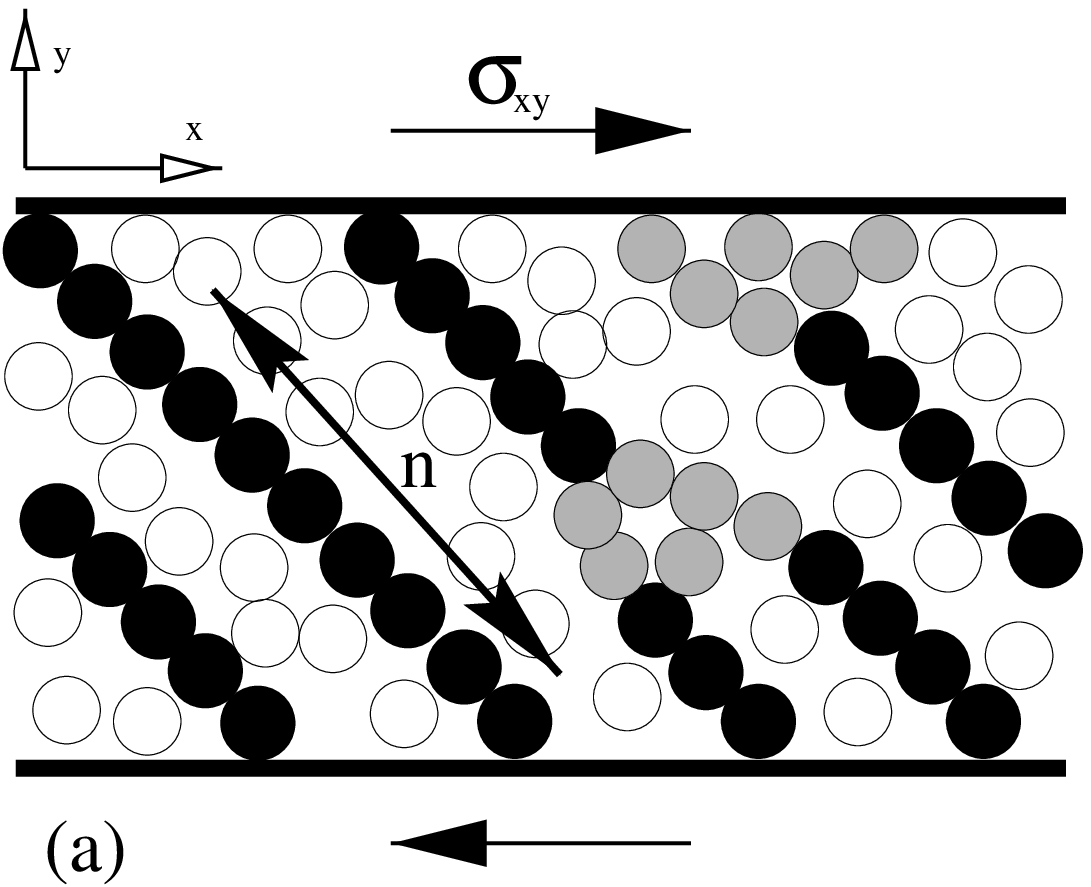}}
\scalebox{.6}{\includegraphics{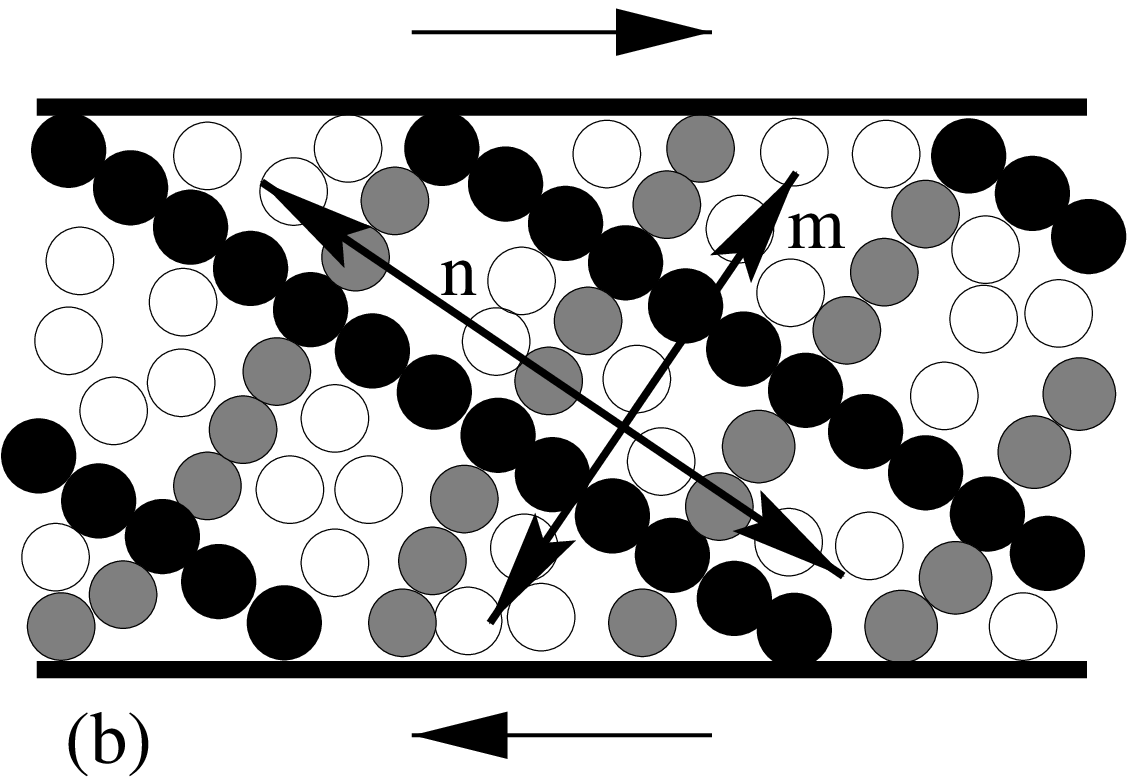}}}
\caption{(a) A jammed colloid (schematic).
Black: force chains; grey: other force-bearing particles; white:
spectators. (b) Idealized rectangular network of force chains.
\label{fig:jammed}}
\end{figure}

Fragile models require that the behaviour of a pile
of cohesionless grains (fragile) is quite different from a hypothetical pile
where each grain is firmly glued to its neighbours on first coming to rest
(elastic). This is plausible because for random packings of glued grains, a
finite fraction of the interparticle forces will be under tension (forbidden
in the cohesionless case) and will remain so until the load is large enough
to cause appreciable elastic deformation of individual grains. For simple
fragile models such as {\sc fpa}, the same criterion defines a crossover between
fragile behaviour and a form of (anisotropic) elastoplasticity.
A fuller discussion of these ideas is given in \cite{PRLroysoc}, where
further references may be found.

\section{Jamming in Colloids: A Simple Model}
Consider a concentrated colloidal suspension of hard particles, confined
between parallel plates at fixed separation, to which a shear stress is
applied. Above a certain threshold of stress, this system exhibits enters a
regime of strong shear thickening; see, e.g. \cite{laun}. The effect can be
observed in the kitchen, by stirring a concentrated suspension of corn-starch
with a spoon. In fact, computer simulations suggest that, at least under
certain idealized conditions, the material will jam completely and cease to
flow, no matter how long the stress is maintained \cite{farrmelrose}. In
these simulations,  jamming apparently occurs because the particles form
``force chains" \cite{dantu} along the compressional direction
(Fig.~\ref{fig:jammed}~(a)). Even for spherical particles the
lubrication films cannot prevent direct interparticle contacts; once
these arise, an array or network of force chains can indeed support the
shear stress indefinitely. (We ignore Brownian motion, here and below,
as do the simulations; this could cause the jammed state to have finite
lifetime.)

\begin{figure}
\centerline{\scalebox{.6}{\includegraphics{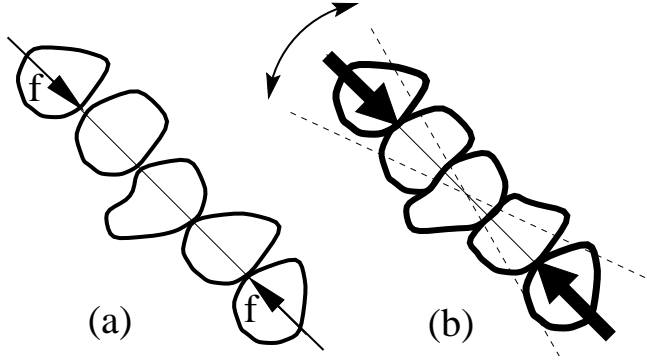}}}
\caption{(a) A force chain of hard particles (any shape) can
statically support only longitudinal compression.
(b) Finite deformability allows small transverse loads
to arise. \label{fig:paths} }
\end{figure}

To model the jammed state, we start from a simple idealization of a force
chain: a linear string of at least three rigid particles in point contact.
Crucially, this chain can only support loads {\em along its own axis}
(Fig.\ref{fig:paths}~(a)): successive contacts must be
collinear, with the forces along the line of contacts, to
prevent torques on particles within the chain \cite{EandO}.
Note that neither friction at the contacts, nor particle aspherity, can
change this ``longitudinal force" rule.

As a minimal model of the jammed colloid, we take an assembly of such force
chains, characterized by a unique director
$\bf n$, in a sea of ``spectator" particles, and incompressible solvent.
This is obviously oversimplified, for we ignore completely any collisions
between chains, the deflections caused by weak interactions with the spectator
particles, and the fact that there must be some spread in the orientation of
the chains themselves \cite{farrmelrose}.
With these assumptions, in static equilibrium, with
no body forces acting, the pressure tensor $p_{ij}$ (defined as $p_{ij}
-\sigma_{ij}$, with $\sigma_{ij}$ the usual stress tensor) must obey
\begin{equation}
p_{ij} = P\delta_{ij} + \Lambda\, n_in_j
\label{nn}
\end{equation}
Here $P$ is an isotropic fluid pressure, and $\Lambda$ ($>0$) a
compressive stress carried by the force chains.

Eq.~(\ref{nn}) defines a material that is mechanically very unusual. It
permits static equilibrium only so long as the applied compression is along
$\bf n$; while this remains true, incremental loads (increase or decrease in
stresses at fixed major compression of the stress tensor) can be
accommodated reversibly, by what is (at the particle contact scale) an elastic
mechanism. But the material is certainly not an ordinary elastic body, for if
instead one tries to shear the sample in a slightly different direction
(causing a rotation of the principal stress axes) static equilibrium cannot be
maintained without changing the director $\bf n$.  Now, $\bf n$ describes the
orientation of a set of force chains that pick their ways through a dense sea
of spectator particles. Accordingly $\bf n$ cannot simply rotate; instead, the
existing force chains must be abandoned and new ones created with a slightly
different orientation. This entails
dissipative, plastic, reorganization, as the particles start to move but then
re-jam in a configuration that can support the new load.

\section{Jamming and Fragile Matter}
Our model jammed colloid is an idealized example of ``fragile
matter": it can statically support applied
shear stresses (within some range), but only by virtue of a
self-organized internal structure, whose mechanical properties have
evolved directly in response to the load itself. Its incremental response can
be elastic only to {\em compatible} loads;
{\em incompatible} loads (in this
case, those of a different compression axis), even if small, will
cause finite, plastic reorganizations. The inability
to elastically support {\em some} infinitesimal loads is our
proposed definition of the term ``fragile" (which, up to now, has not
been given a precise technical meaning in this context).

We argue that jamming may lead {\em
generically} to mechanical fragility, at least in systems with overdamped
internal dynamics. Such a system is likely to arrests as soon as it can
support the external load; since the load is only just supported, one expects
the state to be only marginally stable. Any incompatible perturbations then
force rearrangement; this will leave the system in a newly jammed but (by the
same argument) equally fragile state. This scenario is related, but not
identical, to several other ideas in the literature
\cite{EandO,alexander,soc,kolymbas,balledwards,moukarzel}. These include
the emergence of rigidity  by successive buckling of force chains in
glasses and granular matter
\cite{alexander}; the concepts of self-organized criticality ({\sc soc})
\cite{soc}  (see also \cite{staticavalanche}), and
those of mechanical percolation which underly recent ``hypoplastic"
models of granular matter \cite{kolymbas}. Fragility is also connected with
recent ideas concerning isostaticity and marginal mechanics in (frictionless)
sphere packings \cite{balledwards,moukarzel} (see Section
\ref{marginal}).

Our ideas are, in addition, reminiscent of the (much older) concept of
a self-selecting critical state in theories of soil mechanics
\cite{wood}. However the latter is  primarily concerned with {\em
dilatancy}: the tendency of dense particulate media to expand upon
shearing. Jamming can be viewed as the constant-volume counterpart of
this process: if expansion is prevented, jamming results.

\section{Two Types of Fragility}
Consider the idealized jammed colloid of (Fig.~\ref{fig:jammed}~(a)). So far
we allowed for an external stress field (imposed a the plates) but no body
forces.  What body forces can it now support {\em without} plastic rotation of
the director? Various models are possible.  One is to assume that
Eq.~(\ref{nn}) continues to apply, with
$P({\bf r})$ and $\Lambda({\bf r})$ now varying in space.
If $P$ is a simple fluid pressure, a localized body force can be supported
only if it acts along $\bf n$.
Thus (as in a bulk fluid) no static Green function
exists for a general body force.
(Note that, since Eq.~(\ref{nn}) is already written
as a continuum equation, such a Green function would describes the response to
a load that is localized in space but nonetheless acts on many particles in
some mesoscopic neighbourhood.)

For example, if the particles in Fig.~\ref{fig:jammed}~(a) were to
become subject to a gravitational force along $y$, then the existing force
chains could not sustain this but would reorganize. Applying the longitudinal
force rule, the new shape is easily found to be a catenary, as realized by
Hooke \cite{hooke}, and emphasized by Edwards \cite{EandO}.  On the
other hand,  a general body force can be supported, in three dimensions,
if there are several different orientations of force chain, possibly
forming a network or ``granular skeleton"
\cite{dantu,kolymbas,thornton,radjai}.   A minimal model for this is:
\begin{equation}
p_{ij} = \Lambda_1\, n_in_j +
\Lambda_2\, m_im_j + \Lambda_3\, l_il_j\label{osl}
\end{equation}
with ${\bf n},{\bf m},{\bf l}$ directors along three
nonparallel populations of force chains; the $\Lambda$'s are
compressive pressures acting along these. Body forces cause
$\Lambda_{1,2,3}$ to vary in space.

We can thus distinguish two levels of fragility, according to whether
incompatible loads include localized body forces ({\em bulk} fragility,
{\em e.g.} Eq.~(\ref{nn})), or are limited to forces acting at the boundary
({\em boundary} fragility, {\em e.g.} Eq.~(\ref{osl})). In disordered
systems one should also distinguish between macro-fragile responses
involving changes in the {\em mean} orientation of force chains, and the
micro-fragile responses of individual contacts.
Below we focus on macro-fragility, but in practice the distinction may become
blurred. In any case, these
various types of fragility should not be associated too strongly with minimal
models such as Eqs.~(\ref{nn},\ref{osl}). It is clear that many granular
skeletons
have a complex network structure where many more than three directions of
force chains exist. Such a network may nonetheless be fragile; see Section
\ref{marginal} below.

\section{Fixed Principal Axis ({\sc fpa}) Model}

Returning to the simple model of Eq.~(\ref{osl}), the chosen values of the
three directors (two in 2-d) clearly should depend on how the system came to
be jammed (its ``construction history"). If it
jammed in response to a constant external stress,
switched on suddenly at some earlier time, one
can argue that the history is specified purely by
the stress tensor itself. In this case, if one
director points along the major compression axis 
then by symmetry any others should lie at
rightangles to it (Fig.~\ref{fig:jammed}~(b)).
Applying a similar argument to the intermediate
axis leads to the ansatz that all three directors
lie along principal stress axes; this is perhaps
the simplest model in 3-d. One version of this
argument links force chains with the fabric tensor
\cite{kolymbas}, which is then taken coaxial with the stress \cite{radjai}.

With the ansatz of perpendicular directors as just described,
Eq.~(\ref{osl}) becomes a ``fixed principle axes" ({\sc fpa}) model
\cite{nature,bcccargese}. Although grossly oversimplified, this leads to
nontrivial predictions for the jammed state in the colloidal problem, such as
a constant ratio of the shear and normal stresses when these are varied in the
jammed regime. Such constancy is indeed reported by Laun \cite{laun} in
``the regime of strong shear thickening"; see \cite{PRLroysoc}.

\section{Granular Materials}

We now turn from colloids to granular materials. Although
the formation of dry granular aggregates under gravity is not normally
described in terms of jamming, it is a closely related process. Indeed, the
filling of silos and the motion of a piston in a cylinder of
grains both exhibit jamming and stick-slip phenomena associated with force
chains; see \cite{samchains}. And, just as in a jammed colloid, the
mechanical integrity of a sandpile disappears as soon as the load (in
this case gravity) is removed.

In the granular context, a model like Eq.~(\ref{osl}) is interpreted by saying
that a fragile granular skeleton of force chains is laid down at the time when
particles are first buried at the free surface; so long as subsequent
loadings are compatible with this structure, the skeleton will remain intact.
If in addition the skeleton is rectilinear (perpendicular directors) this
forces the principal axes to maintain forever the orientation they had close
to the free surface ({\sc fpa} model). However, we do not insist on this last
property and other models, based on an oblique family of directors, have also
been developed \cite{nature,PRLroysoc}. 
(The construction history of a sandpile allows this since the orientation
of the free surface and/or gravity provides a reference direction in
addition to that given by the stress tensor at the onset of jamming
\cite{nature}.)
Note also that, for a conical
sandpile we require in addition to Eq.~(\ref{osl}) one further relation
among stresses,  found for example by assuming that $\Lambda_2 =
\Lambda_3$ everywhere; see
\cite{nature}. It turns out that the {\sc fpa} models account quite well for
the forces measured experimentally beneath conical piles of sand, constructed
by pouring cohesionless grains from a point source onto a rough rigid support
\cite{smidhuntley,nature}.
(Note that the two dimensional case -- a wedge -- may have special
features and  the effectiveness of {\sc fpa} for this is much less clear
\cite{PRLroysoc}.) 

As mentioned previously, fragile models such as this show very different
mechanics from conventional forms of elasticity or elastoplasticity. For
example, in 2-d, when combined with stress continuity ($\partial_i\sigma_{ij}
= \rho g_j$ for sand under gravity), Eq.~(\ref{osl}) gives differential
equations for the stress tensor which are hyperbolic
\cite{bcccargese}. With a zero-force boundary condition at the upper
surface of a pile, this gives a well-posed problem: the forces
acting at the base follow uniquely from the body forces by integration.
(Analogous remarks apply to Eq.~(\ref{osl}) in 3-d.) If {\em different} forces
are now imposed at the base, rearrangement is inevitable \cite{evesq}.
(This is boundary-fragile behaviour.) The same does not hold
\cite{PRLroysoc} within a traditional elastoplastic modelling approach
\cite{savagegoddard} whose equations are elliptic in elastic zones. In
such models of the sandpile the forces acting at the base cannot be
found without specifying a displacement field there. To define this
displacement, one would normally invoke as {\em reference state} the one
in which the load (gravity) is removed.  For cohesionless poured sand,
this state is undefined \cite{deGEv},  just as it is for a jammed
colloid which, in the absence of the applied shear  stress, is simply a
fluid.

One route to an elastic reference state is to consider a hypothetical sandpile
where each grain becomes firmly ``glued" to its neighbours (or the base),
upon first coming to rest. The resulting medium is surely elastic, and must
therefore be governed by elliptic equations. This does not mean that it is
a conventional homogeneous elastic continuum (for which the states
of zero strain and zero stress coincide). Indeed, a glued pile built under
gravity will certainly have nonzero stresses if gravity is later removed
\cite{PRLroysoc}. More importantly, for a typical disordered packing of
near-rigid, glued grains, there will arise many {\em tensile contacts} even
under a purely compressive external load. Thus the problem of glued and
unglued piles might, in practice, be extremely dissimilar.

\section{Isostaticity, Marginal Packings, and Fragility}
\label{marginal}
The problem of cohesionless granular media is a
highly nonlinear one, because (a) grains are typically very rigid (on a scale
set by the stresses that arise), and (b) there can be no tensile forces
between any grains in the entire system. The first issue is addressed in
\cite{PRLroysoc} where we show that, for some fragile models, there
is in fact a smooth crossover to more conventional elastoplastic physics when
grains are deformable, with fragility emerging as the limiting behaviour
for rigid particles.

In this rigid particle limit, where the longitudinal force rule ({\sc lfr})
becomes valid, fragility will be recovered in a granular skeleton of force
chains whose coordination number $z=2d$ with $d$ dimension of space (e.g.
Fig.~\ref{fig:jammed}~(b) in two dimensions). This is the same rule as
applies for packings of {\em frictionless} hard spheres, which also obey the
{\sc lfr} -- not because of force chains, but because there is
no friction. Indeed, regular packings of frictionless spheres, which show
``marginal" or ``isostatic" mechanics have been studied in detail recently
\cite{balledwards}.
Note that although our idea of fragility is closely related to the concept of
isostaticity (see e.g. \cite{moukarzel}), it is
apparently not identical; in our models of fragile materials the
isostatic condition applies only to a loadbearing substructure (the force
skeleton) and not the whole packing. 
It would be interesting to test this idea numerically by measuring
the mean coordination number among the subset of particles that carry
strong forces.

Interestingly, Moukarzel has recently
argued \cite{moukarzel} that random packings of frictionless spheres
{\em generically} become isostatic ($z=2d$) in the rigid particle limit.
His arguments
appear to depend only on the {\sc lfr} and the absence of tensile forces, so
they should, if correct, equally apply to any granular skeleton that is made
of force chains of rigid particles. It remains to be
seen whether Moukarzel's (somewhat unintuitive) arguments can be made
rigorous. Pending this, we prefer for the moment our own
qualitative reasoning, which generically links fragility to jamming, as the
motivation for setting $z=2d$ in our simplified models (such as {\sc fpa}) of
the granular skeleton. Such models, in which the skeleton is represented as a
rectilinear or oblique array with no disorder, are obviously convenient for
calculation. However, disorder will not remove the fragility (though it will
change  Eqs.~(\ref{nn},\ref{osl})) unless it causes the mean coordination
number
of the skeleton to increase.

\section{Conclusions}\label{sec:Conclusion}

The jammed state of colloids, if it indeed exists in the laboratory, has not
yet been fully elucidated by experiment. It is interesting the even very
simple models such as Eq.~(\ref{nn}) can lead to nontrivial and testable
predictions (such as the constancy of certain measured stress ratios). Such
models suggest an appealing conceptual link between jamming, force chains, and
fragile matter \cite{PRLroysoc}. However, further experiments are needed
to establish the degree to which they are useful in describing real
colloids.

For granular media, the existence of tenuous force-chain skeletons
is clear \cite{dantu,kolymbas,coppersmith,thornton,radjai}; the question
is whether such skeletons are fragile. Several theoretical arguments
have been given, above and elsewhere, to suggest that this may be the
case, at least in the limit of rigid particles. Moreover, simulations
show strong rearrangement under small changes of compression axis; the
skeleton is indeed ``self-organized" \cite{thornton,radjai}. Experiments
also suggest cascades of  rearrangement \cite{staticavalanche,samchains}
in response to small disturbances. These findings are consistent with
the fragile picture.

The mechanics of fragile models such as Eqs.~(\ref{nn},\ref{osl}) differ
strongly
from those of conventional elasticity. For example, if an infinitesimal
point force is exerted on top of a layer of sand, we expect the resulting
pressure distribution at the base to form an annulus \cite{bcccargese};
an elastic model would predict the maximum pressure to be directly
beneath the applied force. It could be very fruitful to perform new (but
quite simple) experiments of this kind.

\begin{ack}

We thank R. Ball, E.~Clement, S.~Edwards, M. Evans, P.~Evesque, 
P.-G. de Gennes,
G.~Gudehus, J.~Goddard, J. Jenkins, D. Levine, J. Melrose, S. Nagel,
J. Socolar, C.~Thornton, L.~Vanel and T.~A. Witten for discussions.
Work funded in part by EPSRC (UK) GR/K56223 and GR/K76733.
\end{ack}

\end{document}